\documentclass[runningheads]{svmult}
\usepackage{makeidx}   % allows index generation
\usepackage{graphicx}  % standard LaTeX graphics tool
\usepackage{subeqnar}  % subnumbers individual equations
\usepackage{multicol}  % used for the two-column index
\usepackage{physprbb}  % modified textarea for proceedings,
\graphicspath{{fig/}}

\makeindex             % used for the subject index
\usepackage{epstopdf}
\usepackage{amsfonts}   % useful for coding complex math
\usepackage{amsbsy}
\usepackage{amsmath}
\usepackage{natbib}
\bibpunct{(}{)}{;}{a}{}{,}

\newcommand{\beq}{\begin{equation}}
\newcommand{\eeq}{\end{equation}}
\newcommand{\eeqn}[1]{\label{#1}\end{equation}}
\newcommand{\greq}{\begin{equation}\left\{ \begin{array}{l}}
\newcommand{\egreq}{\end{array}\right. \end{equation}}
\newcommand{\Div}{\displaystyle {\rm Div}}
\newcommand{\tio}{\tilde{\omega}}
\newcommand{\na}{ \vec{\nabla} }
\newcommand{\eq}[1]{(\ref{#1})}
\newcommand{\dxi}[1]{\frac{\partial  #1}{\partial \xi}}
\newcommand{\msun}{\mbox{\rm M$_\odot$}}
\newcommand{\er}{\vec{e}_r}
\newcommand{\es}{\vec{e}_s}
\newcommand{\lp}{ \left(}
\newcommand{\rp}{ \right)}
\newcommand{\dr}[1]{\frac{\partial  #1}{\partial r}}
\newcommand{\sth}{ \sin\theta }
\newcommand{\cth}{ \cos\theta }
\newcommand{\dtheta}[1]{\frac{\partial  #1}{\partial \theta}}
\newcommand{\impl}{\; \Longrightarrow\;}
\newcommand{\cst}{{\rm Cst}}
\newcommand{\orou}{ {\qquad {\rm or}\qquad}}
\newcommand{\eps}{\varepsilon}
\newcommand{\lc}{ \left[}
\newcommand{\rc}{ \right]}
\newcommand{\andet}{ \qquad {\rm and}\qquad}
\newcommand{\demi}{\frac{1}{2}}
\newcommand{\tiers}{\frac{1}{3}}
\newcommand{\quart}{\frac{1}{4}}
\newcommand{\YL}{Y^m_\ell}
\newcommand{\od}[1]{\mbox{${\cal O}(#1)$}}
\newcommand{\dz}[1]{\frac{\partial  #1}{\partial z}}
\newcommand{\dtau}[1]{\frac{\partial #1}{\partial \tau}}
\newcommand{\vF}{\vec{F}}
\newcommand{\vg}{\vec{g}}

\begin{document}
\title*{Physical processes leading to surface inhomogeneities: the case
of rotation}
% Relating to Observables of Rapidly\protect\newline 
%Rotating Stars}
%
\toctitle{Physical processes leading to surface
inhomogeneities:\protect\newline the case  of rotation}

%\toctitle{Issues Relating to Observables of Rapidly\protect\newline 
%Rotating Stars}
% allows explicit linebreak for the table of content
%
\titlerunning{Physical processes leading to surface inhomogeneities: the
case  of rotation}

% allows abbreviation of title, if the full title is too long
% to fit in the running head
%
\author{Michel Rieutord}
\authorrunning{Michel Rieutord}

\institute{Universit\'e de Toulouse; UPS-OMP; IRAP; Toulouse, France
\and CNRS; IRAP; 14, avenue Edouard Belin, F-31400 Toulouse, France\\
E-mail: \texttt{mrieutord@irap.omp.eu}}

\date{\today}

\maketitle              % typesets the title of the contribution

\begin{abstract}
In this lecture I discuss the bulk surface heterogeneity of rotating
stars, namely gravity darkening. I especially detail the derivation of
the $\omega$-model of \cite{ELR11}, which gives the gravity darkening
in early-type stars.  I also discuss the problem of deriving gravity
darkening\index{gravity darkening} in stars owning a convective envelope
and in those that are members of a binary system.

\end{abstract}

\section{Introduction}
 
As it has been much discussed in this school, surface inhomogeneities of
stars are more and more frequently detected due to the increasing
sensistivity of the instruments. If correctly understood, and therefore
modeled, these data may open new windows on the interior or on the
history of stars.

The purpose of this lecture is to first briefly review the processes
that lead to such inhomogeneities and then to focus on a very
fundamental one, namely rotation.

In the ancient times, when the eye was the only optical instrument to observe
Nature, the sun was thought as a pure uniform bright disc. The invention
of the telescope by Galileo ruined this idea, showing that the sun was
spherical with spots on it. Presently, it suffices to have a look at
images of the sun taken at short wavelengths to understand that its
surface is certainly not uniform. Such images actually reveal that the
magnetic fields are a prominent cause of this non-uniformity. It looks
like a mess which even impacts the distribution of surface flux and temperature.
A closer look at the magnetic structures but also below, at the
photospheric level, shows that all these heterogeneities evolve with time.
On the photosphere, turbulent convection features the surface with two
important scales: granulation and supergranulation \cite[][]{RR10}. Even
the bulk surface rotation is not uniform. This differential rotation,
known since the nineteenth century, with fast equatorial regions and
slow polar regions, is now understood as driven by Reynolds stresses
coming from the turbulence in the solar convection zone.

Hence, the surface of the sun teaches us that we should expect
non-uniform velocities, temperatures, flux and magnetic fields at the surface
of all low-mass stars. But one consequence of the strong mixing imposed
by turbulent convection and the ever changing magnetic fields is that
the solar photosphere has a uniform chemical composition!

But uniform chemical composition is certainly not possible when turbulent
convection disappears and no longer mixes the surface layers, that is when
we consider stars of higher mass with an outer radiative envelope. There,
combination of magnetic fields with microscopic diffusion processes
(gravitational settling or radiative acceleration) may
on the contrary lead to chemical surface inhomogeneities
\cite[][]{VV82,alecian13,korhonen_etal13}. But even when magnetic fields
are absent at their surface, early-type stars are still endowed with
a non-uniform surface: absence of magnetic field is indeed correlated
with fast rotation, a feature that makes the polar caps brighter than
the equatorial regions.

From the foregoing presentation we see that three processes, intrinsic to
each star may lead to surface inhomogeneities: rotation, convection and
magnetic fields. There is a fourth one, but extrinsic to the star
itself, namely binarity. A companion indeed raises tides, illuminates
one side of the star or may even transfer mass.

Within these four physical processes that make the surface of stars not
uniform, we shall concentrate on the most simple, namely rotation,
which, a priori leads surface variations that only depend on the
latitude. We shall discuss in detail this very basic physical effect,
leaning on the recent works of \cite{ELR11,ELR12}.

\section{The energy flux in radiative envelopes of rotating stars}
\index{radiative envelope}\index{energy flux}

\subsection{von Zeipel 1924}\index{von Zeipel 1924}

In a seminal paper, \cite{vz24} showed that a rotating star may be
brighter at the poles than at the equator. This result is quite simple
to derive if we assume that the star is barotropic and that the energy
flux is given by Fourier's law. Indeed, if the star is barotropic,
meaning that its equation of state can be simplied to 

\[ P\equiv P(\rho),\]
it implies that there exist a hydrostatic solution in the rotating
frame and that all thermodynamic quantities can be expressed as a
function of the total potential $\Phi$ (i.e. gravitational plus
centrifugal). Hence, one writes

\[ \rho\equiv\rho(\Phi), \quad T\equiv T(\Phi),\quad {\rm etc.} \]
Then, using Fourier's law to derive the heat flux, one finds

\[ \vF_{\rm rad} = -\chi\na T = -\chi(\Phi)T'(\Phi)\na\Phi =
K(\Phi)\vg_{\rm eff}\]
whence von Zeipel law

\[ T_{\rm eff} = K g_{\rm eff}^{1/4}\qquad {\rm on\; the\; surface}\;
\Phi=\cst \]
This result is simple but incorrect. Indeed, barotropic stars are
realized in two cases: either the star is isentropic and thus fully
convective or it is isothermal but this can hardly be the
case\footnote{Some models of prestellar core use this hypothesis,
sometimes.}. So the closest case that may match the barotropic state is
that of a fully convective star, but in such a case the flux cannot
be derived from the Fourier's law. In fact, these hypothesis
(barotropicity and heat diffusion) lead to a contradiction. Indeed, we
may note that the total potential and the temperature would verify in
the envelope of the star,

\greq
\Div(\chi\na T) = 0 \\
\Delta \Phi = 4\pi G\rho +2\Omega^2
\egreq
which leads to

\[ \Div(\chi(\Phi)T'(\Phi)\na\Phi)=0 \qquad \Longleftrightarrow\qquad  4\pi
G\rho+2\Omega^2+(\ln(\chi T'))'g_{\rm eff}^2 = 0\]
On an equipotential, $\rho$ is constant as well as $(\ln(\chi T'))'$,
but $g_{\rm eff}$ is not constant. Hence, the latter equation is
impossible. The reason for that is that for a rotating star where
heat is transported by diffusion, a barotropic state cannot be and
should be replaced by a baroclinic state.\index{baroclinic!state} In
such a state, isobars, isotherms or equipotential are all different,
not very different, but different. This is the normal state that comes
from the fact (basically) that temperature, pressure, and gravitational
potential all obeys different and independent equations. The barotropic
state is therefore rather peculiar \cite[but see][for a more detailed
presentation]{R06b}.

Now one may wonder if it is possible to derive the dependence of the flux
with latitude for a rotating star without computing the whole stellar
structure and the associated flows as in \cite{ELR13}.  Fortunately,
this is indeed possible as \cite{ELR11} have shown.  It is not as simple
as von Zeipel law, but it has the merit of relying on
controllable hypothesis.

\subsection{The idea of the $\omega$-model}\index{$\omega$-model}

In the following we shall first restrict ourselves to the case of
early-type stars, that is to stars that have a radiative envelope around
a convective core. We'll discuss the case of convective envelope in the
next section.

Within the envelope of a star the flux just obeys:

\beq \Div\vF = 0\eeqn{fluxeq}
namely energy is conserved and there are no energy sources.

This is a single equation, not enough to determine the two components,
($F_r, F_\theta$), of the flux, but if we add a constraint to the flux
we may find it. We thus assume that the flux is anti-parallel to the
effective gravity

\beq \vF = -f(r,\theta)\vg_{\rm eff}\eeqn{hypo}
In order to avoid an additional unknown, we shall take the effective
gravity $\vg_{\rm eff}$ as given by the Roche model. In such a case we
shall see that the flux function $f$ can be determined and that the
latitude variation of the flux depends on a single parameter $\omega$
defined as the ratio of the angular velocity to the keplerian angular
velocity at the equator. In other words, the flux depends on

\beq \omega = \frac{\Omega}{\Omega_k} =
\Omega\lp\sqrt{\frac{GM}{R_e^3}}\rp^{\!\!-1} \; .
\eeqn{omega}
Thus, we shall call this model the {\em $\omega-model$} to emphasize the
crucial role of the reduced angular velocity $\omega$.

However, before going any further, we may wonder whether the assumptions
are strong or not, especially \eq{hypo}.

\begin{figure}[t]
\centerline{\includegraphics[width=0.9\linewidth,angle=0]{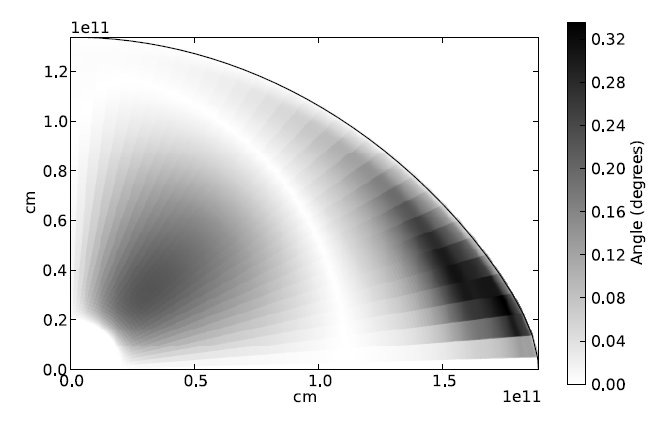}}
\caption[]{Misalignment between pressure gradient and flux for a
configuration with a flatness $\sim30$\% \cite[from][]{ELR11}.}
\label{align}
\end{figure}

In a radiative zone, the configuration is baroclinic so vectors are
surely not aligned but fortunately we can now revert to 2D-models
to get an idea of the misalignment. As shown in Fig.~\ref{align},
the misalignement remains small, less than a degree, even if the star
rotates close to criticality.

Thus, even if the envelope is the seat of baroclinic flows,
\index{baroclinic!flows} the
misalignment is small. Actually, the baroclinic torque $(\na
P\times\na\rho)/\rho^2$ does not need a strong misalignment of the
vectors to be efficient at driving baroclinic flows because the two
gradients (of pressure and density) are already quite strong.

Let us pursue somewhat. From \eq{fluxeq} and \eq{hypo}, we have

\[ \Div\vF = 0 \quad \Longleftrightarrow \quad \Div(f\na\Phi) =0\]
thus

\beq \vg_{\rm eff}\cdot\na\ln f = -2\Omega^2\eeqn{eqf}
because $\Delta\Phi=-2\Omega^2$; hence,

\beq \dxi{\ln f} = -\frac{2\Omega^2}{g_{\rm eff}}\eeqn{fhvar}
where we introduced the local vertical coordinate
$\xi$. Equality \eq{fhvar} shows that $\dxi{\ln f}$, and therefore $f$,
has latitudinal variations similar  to those of $g_{\rm eff}$. Hence the
horizontal variations of the flux cannot be given by von Zeipel law. In
other words $T_{\rm eff}/g_{\rm eff}^{1/4}$ cannot be constant.

\begin{figure}[t]
\centerline{\includegraphics[width=0.45\linewidth,angle=0]{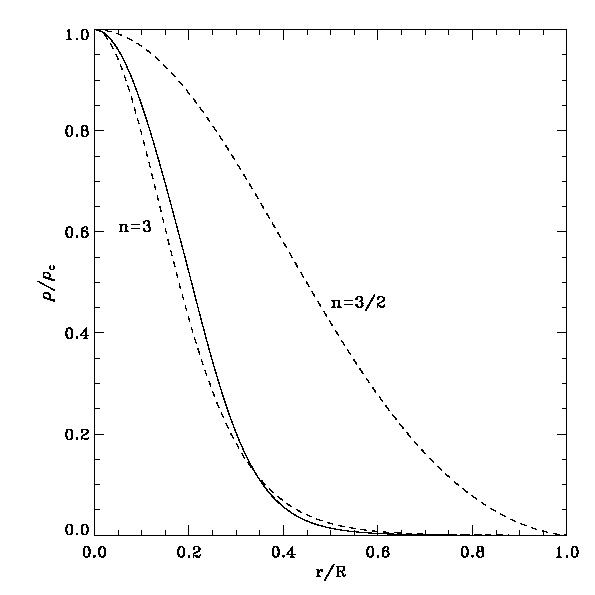}
\includegraphics[width=0.45\linewidth,angle=0]{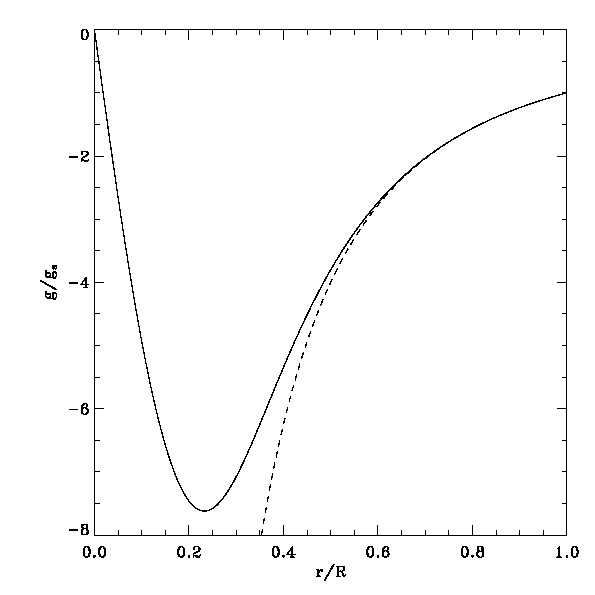}}
\caption[]{Left: Density profile of a M=5~\msun\ ZAMS, non rotating star
(solid line) together with that of a n=3/2 and n=3 polytropes. Right:
The interior gravity of the same stellar model (solid line) together
with the $-1/r^2$ Roche model (dashed line). The ZAMS model is an ESTER
model \cite[e.g.][]{ELR13}}
\label{poly}
\end{figure}

The second hypothesis is the use of the Roche model. This model assumes
that the whole mass of the star is concentrated at the centre thus
leading to a gravitational potential in $1/r^2$ everywhere. For the
regions we are interested in, namely the envelope of early-type stars,
this is a rather good approximation since these stars are usually said
to be ``centrally condensed''. In Fig.~\ref{poly} we show the density
profile of a non-rotating 5~\msun\ star along with the profile of two
polytropes. We see that the n=3 polytrope represents fairly well the
density profile of the star and that the n=3/2-polytrope, which is a
very good model for fully convective stars, is much less ``centrally
condensed". Hence, gravity in the outer envelope of an early-type star
is well represented by the Roche model (see Fig.~\ref{poly} right). The
interior discrepancy with the Roche model has no consequence for our
purpose.

\subsection{The derivation of $f(r,\theta)$}

$f$ is given by \eq{eqf} but we first need to scale this function so as
to introduce a non-dimensional function $F$ that accounts for the radial
and latitudinal variation of the flux. This is easily done if we observe
that near the star's centre

\[ \vF \sim \frac{L}{4\pi r^2}\er \qquad \vg_{\rm eff}\sim-\frac{GM}{r^2}\er\]
So that we may set

\beq f(r,\theta) = \frac{L}{4\pi GM}F(r,\theta)\eeqn{Fdef}
with

\[ \lim_{r\rightarrow0} F(r,\theta) =1\]
Then, we scale the gravity with $GM/R_e^2$ and the length scale with the
equatorial radius $R_e$. The scaled angular velocity is therefore given by

\[ \omega = \frac{\Omega}{\Omega_k} = \Omega\lp\sqrt{\frac{GM}{R_e^3}}\rp^{\!\!-1} \]
At this point we should underline that the angular velocity is scaled
by the keplerian angular velocity given by the equatorial radius. It is
often the case in the literature that the scale of angular velocity is the
critical velocity associated with the Roche model of the considered mass
$M$ \cite[e.g.][for instance]{monnier_etal12}. This gives a different
$\omega$ (i.e. fraction of critical velocity). We give in appendix the
relation between these two ways of appreciating angular velocity.

We now proceed to the derivation of $F$ the scaled version of $f$. From
\eq{Fdef} and \eq{eqf} we get

\[ \lp \frac{1}{\omega^2r^2} - r\sin^2\theta\rp\dr{F} -
\sth\cth\dtheta{F} = 2F\]
With $F(0,\theta)=1$ we have all the elements for solving the equation
for $F$.

First, we solve for $\ln F$, namely,

\beq \lp \frac{1}{\omega^2r^2} - r\sin^2\theta\rp\dr{\ln F} -
\sth\cth\dtheta{\ln F} = 2\; .\eeqn{eqF2}
If we set $\ln F= \ln G + A(\theta)$, so that $A(\theta)$ removes the
RHS of \eq{eqF2}, we immediately find that

\[ A'(\theta) = -2/\sth\cth \impl A(\theta) = -\ln(\tan^2\theta)\; .\]
But we still have to solve the homogeneous equation, namely

\beq  \lp \frac{1}{\omega^2r^2} - r\sin^2\theta\rp\dr{\ln G} -
\sth\cth\dtheta{\ln G} = 0\eeqn{eqG}
Such a first order partial differential equations is solved by the
method of characteristics. We therefore look for places where $\ln G$
is constant. These places are called the characteristics curves of
$G$. They are such that

\[ \dr{\ln G}dr + \dtheta{\ln G}d\theta = 0\]
but $G$ also verifies \eq{eqG} so that we can eliminate
$\dr{\ln G}$ and $\dtheta{\ln G}$ and get

\beq \lp \frac{1}{\omega^2r^2} - r\sin^2\theta\rp d\theta + \sth\cth dr
= 0\eeq
which is the equation of characteristics.

We first observe that we may multiply this equation
by any function $H(r,\theta)$ without changing anything. So we may
also look for $h$ such that

\greq
\displaystyle{\dr{h} = H\sth\cth}\\
\\
\displaystyle{\dtheta{h} =H\lp \frac{1}{\omega^2r^2} - r\sin^2\theta\rp}
\egreq
where $H$ needs to be chosen so that this system can be integrated. After
trial and error, we find that $H=\omega^2r^2\cth\cot\theta$ is the right
function. Thus

\greq
\displaystyle{\dr{h} = \omega^2r^2\cos^3\theta}\\
\\
\displaystyle{\dtheta{h} =\frac{\cos^2\theta}{\sth}-\omega^2r^3\cos^2\theta\sth}
\egreq
and the solution is

\[ h(r,\theta)  = \tiers\omega^2r^3\cos^3\theta+\cth+\ln\tan(\theta/2)\]
The curves $h(r,\theta)=\cst$ are the characteristics. Note that the
polar equation of a characteristic, namely the dependence $r\equiv
r(\theta)$, is just implicitly known, and depends on the chosen constant.

Now, we know that $\ln G$ or $G$ is constant on the curves where
$h(r,\theta)=\cst$. So we can write

\beq G\equiv G(h) \; .\eeq
It means that the variations of $G$ with $(r,\theta)$ are through those of
$h(r,\theta)$ only. So we find that

\[ \ln F = \ln G(h) - \ln\tan^2\theta \orou F = \frac{G(h(r,\theta))}{\tan^2\theta}\]
This is the solution of the partial differential equation, but
it is up to an arbitrary function $G(h)$ that we should determine. For
that, we need to revert to the boundary conditions, namely that
$F(0,\theta)=1$. We thus need to impose

\beq \frac{G(h(0,\theta))}{\tan^2\theta}=1\eeq
or

\beq G(\cth+\ln\tan(\theta/2)) = \tan^2\theta\eeq
for all $\theta$.
This is certainly a weird expression of $G$, but actually it is sufficient.
Let's introduce the function $h_0$ such that

\beq h_0(\theta) = \cth+\ln\tan(\theta/2)\eeq
Hence, we have

\[ (G\circ h_0)(\theta) = \tan^2\theta\]
or

\[ G\circ h_0 = \tan^2 \qquad \impl\quad G=\tan^2\circ h_0^{-1}\]
so formally, the solution for $G$ is

\[ G(r,\theta) = \tan^2(h_0^{-1}(h(r,\theta)))\]

To make it more understandable, we set

\beq \psi = h_0^{-1}(h(r,\theta))\eeq
so that
\[ h_0(\psi) = \tiers\omega^2r^3\cos^3\theta+\cth+\ln\tan(\theta/2)\]
or

\beq \cos\psi+\ln\tan(\psi/2) =
\tiers\omega^2r^3\cos^3\theta+\cth+\ln\tan(\theta/2)\eeqn{eqpsi}
which is a transcendental equation for $\psi$. However, it is not
difficult to solve numerically (we know that when $r$ or $\omega$
are small $\psi\simeq\theta$). So finally we find

\beq F(r,\theta) = \frac{\tan^2(\psi(r,\theta))}{\tan^2\theta}\eeqn{solF}
where $\psi(r,\theta)$ is given by \eq{eqpsi}.

\subsection{Two interesting latitudes}

$F$ seems to be singular at the pole ($\theta=0$) and at the equator
($\theta=\pi/2$). Let us explore these two latitudes.

Starting with the pole, we see that if $\theta\ll1$, then, from
\eq{eqpsi}, we find that $\psi\ll1$ as well. Indeed, for small values
of the angles we have

\[ 1+\ln\tan(\psi/2) \simeq
\tiers\omega^2r^3 +1+\ln\tan(\theta/2)\]
so that

\beq \psi\simeq \theta e^{\omega^2r^3/3}\eeq
and

\beq F(r,0) = e^{2\omega^2r^3/3}\eeq
which gives the values of $F$ along the rotation axis.

The equator is more complicated. We need to know that if $\eps\ll1$ then

\[ \ln\lp\tan\lc\frac{\pi}{4}-\eps\rp\rc =
-\eps-\frac{1}{6}\eps^3-\cdots\]
With this asymptotic expansion we find

\[ F(r,\pi/2) = (1-\omega^2r^3)^{-2/3}\]

\subsection{The final solution of the $\omega$-model}

Back to the definitions we started with, we can express the flux with
the effective gravity in the following way:

\beq \vF = -\frac{L}{4\pi GM}F(\omega,r,\theta)\vg_{\rm eff}\eeq
so that we also get the effective temperature

\beq T_{\rm eff} = \lp\frac{L}{4\pi\sigma
GM}\rp^{1/4}\sqrt{\frac{\tan\psi}{\tan\theta}}\; g_{\rm eff}^{1/4}\eeq
From this expression, we see that the function
$\sqrt{\tan\psi/\tan\theta}$ shows the deviation from the von Zeipel law.

Noting that

\[ \vg_{\rm eff} = \frac{GM}{R_e^2}\lp-\frac{\er}{r^2} +
\omega^2r\sth\es \rp\]
for the Roche model ($\es$ is the unit radial vector of cylindrical
coordinates and $\er$ that of spherical coordinates). We find that the
flux is given by

\beq \vF = -\frac{L}{4\pi R_e^2}\lp-\frac{\er}{r^2} +
\omega^2r\sth\es \rp F(\omega,r,\theta)
\eeq
which shows that it depends only on $\omega$ and a scaling factor
$\frac{L}{4\pi R_e^2}$, hence the name ``$\omega$-model".

\subsection{Comparison with 2D models: a test of the $\beta$- and
$\omega$- models}

\begin{figure}[t]
\centerline{\includegraphics[width=0.5\linewidth,angle=0]{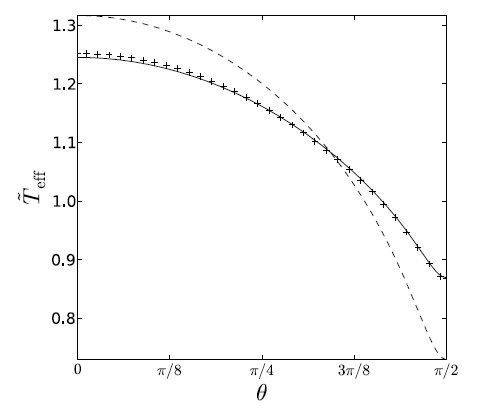}
\includegraphics[width=0.5\linewidth,angle=0]{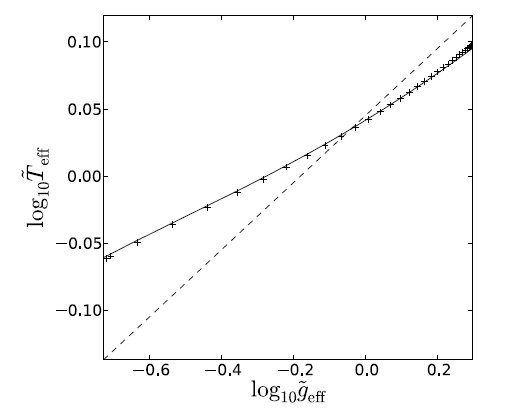}}
\caption[]{Left: Scaled effective temperature as a function of colatitude
for a 3 M$_\odot$ model at $\Omega=0.9\Omega_k$. The solid line shows
the prediction of the simplified model, 'pluses' show the prediction of
a fully 2D ESTER model including differential rotation \cite[see][for
details]{ELR13}, while the dashed line is for the von Zeipel law. Right:
With the same symbols as on left, the effective temperature as a function
of the effective gravity \cite[plots from][]{ELR11}.}
\label{figteff}
\end{figure}

\begin{figure}[t]
\centerline{\includegraphics[width=0.7\linewidth,angle=0]{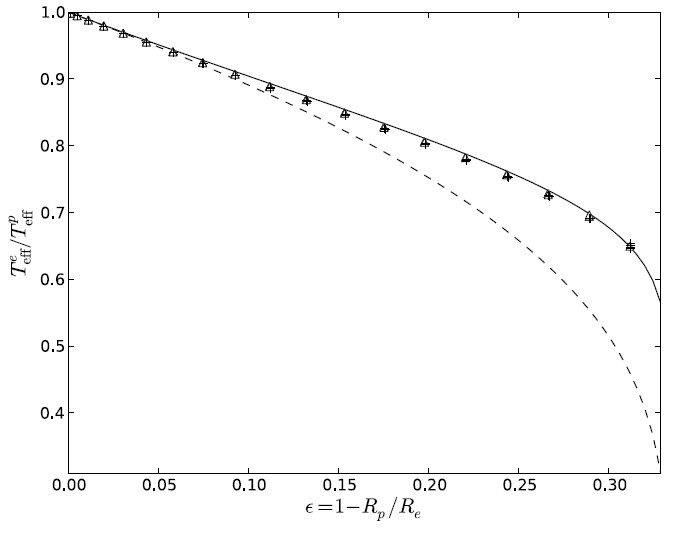}}
\caption[]{Variation of the ratio of effective temperature at pole and
equator as a function of the flatness of the star. Symbols are the same
as in Fig.~\ref{figteff} \cite[plots from][]{ELR11}.}
\label{figcontrast}
\end{figure}

After the foregoing mathematical developments we certainly would like
to compare the results of this modeling to more elaborated models. For
this purpose, we compared the latitude variations of the flux with the
prediction of fully two-dimensional ESTER models \cite[][]{ELR13}. We
recall that ESTER models give a full solution of the internal structure
of a rotating early-type star including the differential rotation
and the meridional circulation driven by the baroclinicity of the
envelope. They also include the full microphysics (opacity and equation
of state) from OPAL tables.  Fig.~\ref{figteff} and \ref{figcontrast}
show that the $\omega$-model matches very well the output of the full
ESTER models. Moreover, we also note that the dependence of the effective
temperature versus gravity is close to but not exactly a power law.

Observational data often show the polar-equator contrast in effective
temperature in terms of the exponent $\beta$ defined as

\beq T_{\rm eff} \propto g_{\rm eff}^\beta \eeqn{betadef}
We shall call this approximate modeling the ``$\beta$-model". Actually,
note that \eq{betadef} demands that

\beq \beta = \left.\frac{\partial\ln T_{\rm eff}}{\partial\ln g_{\rm
eff}}\right|_{r=R(\theta)}\eeqn{betaloc}
where $R(\theta)$ is the radius of the star at colatitude $\theta$. Since
the relation between $T_{\rm eff}$ and $g_{\rm eff}$ is not a power law,
$\beta$ is not constant on the surface of a rotating star. It varies
between two extreme values that we can also compute.

To make things simpler we therefore define the $b$-exponent as follows:

\beq T_e=T_p\lp\frac{g_e}{g_p}\rp^{\! b} \orou b =
\frac{\ln(T_e/T_p)}{\ln(g_e/g_p)}\eeqn{betaglob}
where the indices $e$ and $p$ refer to the equator and pole
respectively. $T$ and $g$ designate the effective temperature and
effective surface gravity.

From the polar and equatorial expression of the flux, we get

\[ F_e=(1-\omega^2)^{-2/3}g_e \andet F_p=e^{2\omega^2r_p^3/3}g_p\]
for the $\omega$-model, while, from the Roche model,

\[ \frac{g_e}{g_p} = r_p^2(1-\omega^2) \quad {\rm with}\quad
r_p=\frac{1}{1+\omega^2/2}\]
where $r_p$ is the polar radius. So we find

\[ \lp\frac{T_e}{T_p}\rp^4 = \frac{(1-\omega^2)^{1/3}}{(1+\omega^2/2)^2}
e^{-2\omega^2r_p^3/3}\]
and

\beq b = \quart -\frac{1}{6}\frac{\ln(1-\omega^2) +
\omega^2r_p^3}{\ln(1-\omega^2) -2\ln(1+\omega^2/2)}\eeqn{bexp}

\begin{figure}[t]
\centerline{\includegraphics[width=0.9\linewidth,angle=0]{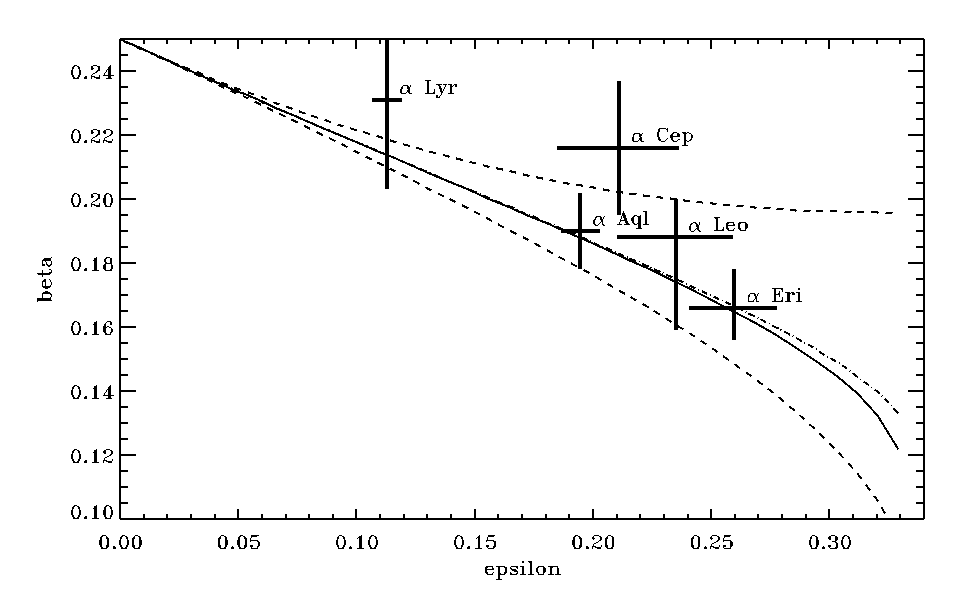}}
\caption[]{The $\beta$-values from various models: The solid line shows the
$b$-exponent of the $\omega$-model, while the dot-dashed line shows the
corresponding ESTER model. The extra dashed lines give the range of
$\beta$ values spawned at the stellar surface by a $\beta$-model. Data
from interferometric observations of some early-type stars are shown
\cite[from][]{domiciano_etal14}.}
\label{betamod}
\end{figure}

We  plotted in Fig.~\ref{betamod} the values of $b$ with increasing
values of the flatness (namely with increasing
rotation). In this figure we see that the $b$-exponent is close to
a linear dependence $b = \quart
-\frac{1}{3}\eps$ up to $\eps=0.3$. However, note that
since the true dependence is not a power law, $\beta$, as given by
\eq{betaloc}, varies at the surface of a given star. We also show in
Fig.~\ref{betamod} its range of variation and it is clearly not
negligible when $\eps$ is larger than $\sim0.15$. This means that if we
had access to a very high spatial resolution of the stellar surface we
would find different $\beta$'s whether we look at the pole (large
values) or at the equator (low values).

If the $\beta$-model is a poor representation of the latitudinal
variation of the flux, can we devise a better one? Surely, a
decomposition of the effective temperature on the spherical harmonics
basis, namely

\[ T_e(\theta) = \sum_{l,m} t^l_m\YL\]
has the advantage of being model independent. The coefficients of the
expansion are the results of observations. Such an expansion is already
used for the description of spotted stars for the reconstruction of
their magnetic fields (e.g. \citealt{donati_etal06} but see also the
lecture of O.  Kochukhov in this volume).

In Fig.~\ref{betamod}, we also show the observationally derived values for
a few early-type stars observed with interferometers. The matching is quite
remarkable, even if some cases like $\alpha$ Cep certainly need a more detailed
study.

%\begin{figure}[t]
%\centerline{\includegraphics[width=0.5\linewidth,angle=0]{fig/beta_obs.jpg}}
%\caption[]{Observationally derived $\beta$-values for various stars
%(Domiciano et al. 2014).}
%\label{betaobs}
%\end{figure}

To finish with the case of early-type stars, let us consider the case
of small rotations. We may first derive the linear dependence of the
$b$-exponent with $\eps$. From \eq{bexp} we get

\beq b = \quart -\frac{1}{6}\omega^2 +\od{\omega^4} \qquad {\rm
or}\qquad b = \quart -\frac{1}{3}\eps + \od{\eps^2}\eeq
where we observed that $\eps=1-r_p$. This expression shows that in the
limit of small rotation we recover von Zeipel law. To understand the origin
of this property, it is useful to reconsider the $\omega$-model and the
expression of the function $F(r,\theta)$. Let us first solve \eq{eqpsi}
in the limit $\omega\ll1$. This yields

\beq \psi = \theta +\tiers \omega^2\sth\cth+\od{\omega^4}\eeq
From this relation, we derive the asymptotic expression of $F(r,\theta)$
at low $\omega$, namely

\[ F(r,\theta) = 1+ \frac{2}{3}\omega^2r^3\]
The latitudinal dependence has disappeared. Hence the latitudinal
variations of the flux are those of the effective gravity. Therefore, von
Zeipel law applies at low rotation rates. We can understand this result,
if we recall that in the limit of zero rotation, the star is spherical
and all surfaces of constant pressure, temperature, etc. are spheres so
that we can consider the gravitational potential or the pressure as the
independent variable.  Thus we recover a kind of barotropic situation
where one can use a relation between pressure and density, and derive a
von Zeipel law.

\section{The case of convective envelopes}\index{convective envelopes}

\subsection{Lucy's problem}\index{Lucy's problem}

In the sixties it was realized that gravity darkening was very important for
the interpretation of light curves of contact binaries (like the W
UMa-type stars). But most of these stars are low-mass stars, thus with a
convective envelope. The use of von Zeipel law, which is based on heat
diffusion, was therefore doubtful.

So \cite{lucy67} asked: { ``What is the gravity-darkening law  appropriate
for late-type stars whose subphotospheric layers are convective?"}
Lucy's reasoning was the following.

\begin{figure}[t]
\centerline{\includegraphics[width=0.9\linewidth,angle=0]{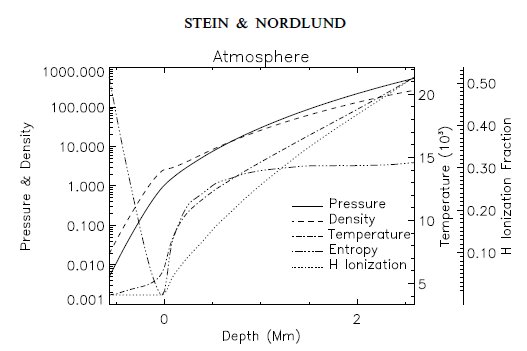}}
\caption[]{Thermodynamic profile of the Sun according to Stein \&
Nordlund 1998.}
\label{entropyjump}
\end{figure}

In the convective envelope of a rotating star, if we go deep enough, we
should reach a medium of constant entropy. This value should be the same
whatever the latitude. 1D models show that the entropy jumps from a
minimum near the surface (where the convective driving ceases)
to a plateau in the deep layers where convective mixing is efficient
(see Fig.~\ref{entropyjump}). Lucy argues that the value of the entropy
$s$ on this plateau is a function of the surface gravity $g_s$ and
effective temperature $T_{\rm eff}$. He thus writes

\beq s\equiv s(g_s,T_e) \eeq
In the case of a rotating star, where $g_s$ and $T_{\rm eff}$ vary, we
must have

\beq s(g_s,T_e)=s_0 \eeq
where $s_0$ is the entropy on the plateau. If we differentiate this
expression with respect to $g_s$ and $T_e$, we find that

\[ \frac{\partial s}{\partial g_s}dg_s + \frac{\partial s}{\partial T_e}dT_e =
0\]
in the deep layers of the rotating star. Since we admit that $T_e \propto
g_s^\beta$, then we have

\beq \frac{\partial s}{\partial \ln g_s} +\beta\frac{\partial s}{\partial
\ln T_e} =0\eeqn{betalucy}
Thus, if we are able to evaluate the values of the above partial derivatives,
we can obtain $\beta$. For that, Lucy considered various 1D neighbouring
models (we do not know how the variations were made), and evaluated
the partial derivatives so as to find $\beta$. Using five stellar models (3 with
M=1\msun, 2 with M=1.26\msun), he found that

\[ 0.069 \leq \beta \leq 0.088\]
Lucy adopted $\beta=0.08$ as a representative value.

\subsection{A new derivation of Lucy's result}

It is interesting to note that Lucy's results may be derived from simple
considerations on one dimensional stellar models in the solar mass
range. 

Let us first recall that the surface of a star is usually determined by
a surface pressure given by

\beq P=\frac{2g_s}{3\kappa}\eeqn{bcp}
where $g_s$ is the surface gravity and $\kappa$ an average opacity. This
boundary condition comes from the assumption of hydrostatic equilibrium
of the atmosphere, namely

\[ \dz{P} = -\rho g\quad \Longleftrightarrow \quad\frac{1}{\rho\kappa}\dz{P} =
-\frac{g}{\kappa}\quad \Longleftrightarrow
\quad\dtau{P} = \frac{g}{\kappa}\]
where the last relation is integrated from the zero optical depth down
to $\tau=2/3$. In the range of density and temperatures typical of the
solar type stars, opacity may be approximated by a power law of the form:

\beq \kappa = \kappa_0 \rho^\mu T^{-s}\eeqn{opapw}
For instance Christensen-Dalsgaard uses
$\mu=0.408$ and $s=-9.283$ for the sun \cite[e.g.][]{CDGR95}.

Now in convective envelopes, the variation of pressure and density
are related to temperature through

\[ P\propto T^{n+1} \andet \rho\propto T^n\; .\]
namely with a polytropic law with $n=3/2$.

Using the foregoing power laws for the opacity, pressure and density,
we can express gravity as a function of temperature. We find that

\[ g \propto T^{n(\mu+1)+1-s}\]
Identifying temperature and effective temperature, we find a gravity
darkening exponent which reads:

\beq \beta= \frac{1}{n(\mu+1)+1-s}\eeqn{beta}
Using Christensen-Dalsgaard's solar values and $n=3/2$, the foregoing
expression yields

\[ \beta \simeq 0.0807\]
which is precisely the value found by Lucy. This is no surprise since
Lucy used models similar to solar models, so the power law fit of
Christensen-Dalsgaard is appropriate.

This derivation clearly shows that this $\beta$-exponent, as defined
by \eq{betalucy}, depends on the chemical properties of the surface through
the opacities.

\subsection{Can Lucy's law represent a gravity darkening effect?}
\index{gravity darkening}

The foregoing derivation of Lucy's result enlights us on the origin of
Lucy's value of the $\beta$ exponent. We see that it is essentially due
to the strong dependence of opacity with temperature in the surface
layers. Since the values for $\mu$ and $s$ are chosen to fit the table
values in some range of density and temperature, we understand that
Lucy's result applies only to stars similar to the Sun, in terms of
gravity and effective temperature. We may note, as \cite{ELR12}, that if
the opacity law extend in the deep layers so as to control the structure
of the envelope and leads to a radiative one, then $\beta=1/4$ because
the polytropic index is $n=(s+3)/(\mu+1)$. We recover the previous
result for non rotating radiative envelopes. We see that when the
opacity is such that the polytropic index is less than 3/2, and the
envelope is convective, the $\beta$ is governed by the opacity of the
surface layers, those which are assumed to be transparent and fixing the
atmosphere. The structure of the envelope is close to the adiabatic
index n=3/2.

Now we wish considering the case of rotating stars. The question is
whether we can use the foregoing value of the exponent, if we consider
a fast rotating star of solar type. A first obstacle is the validity
of the boundary condition \eq{bcp}, which relies on a hydrostatic
equilibrium. When rotation is present such an equilibrium is impossible
because of baroclinicity (for the same reason as the origin of the
so-called von Zeipel paradox, see \citealt{R06b}). The proper boundary
condition, replacing \eq{bcp} should be derived from

\[ \vec{v}\cdot\na\vec{v} = -\frac{1}{\rho}\na P -\na\Phi\]
where $\vec{v}$ is the fluid velocity in an inertial frame. Basically the
flow is a differential rotation plus some weak meridional currents. The
important point is that the differential rotation is latitude dependent.
Hence, if we were to use some pressure boundary condition like \eq{bcp},
we should expect some extra variations from this latitudinal
differential rotation. But this is likely not the whole story as we shall
discuss it now.

If we consider the deep convective envelope of a rapidly rotating star,
we might consider too contradicting effects. First the Coriolis effect:
analysis of a linear stability of a convectively unstable layer shows
that polar regions are less unstable than equatorial ones. This a
consequence of the presence of the Coriolis force. This force indeed
prevents variations of the velocity field along the rotation axis (the
so-called Taylor-Proudman theorem). It shows up in numerical simulations
as convective rolls parallel to the rotation axis near the equatorial
regions \cite[][]{GO93}. For stars this may imply that the convective flux
is larger in the equatorial plane than in the polar region, thus meaning a
negative $\beta$. However, in the equatorial plane the effective gravity
is less, and so is the buoyancy force. This is the effect of centrifugal
force, which therefore points to more flux in the polar region (thus for
a positive $\beta$). The conclusion of the foregoing argument is that
nothing is clear. We may only guess that if Lucy's law applies, it is
for slowly rotating stars of solar type only. This is a deceptive conclusion
since we may have interesting data only on fast rotators or evolved stars
with weak self-gravity. In addition, the previous remarks do not mention
the magnetic fields that are almost unavoidable in late-type stars.

%\begin{figure}
%\centerline{\includegraphics[height=5cm]{fig/schem.jpg}}
%\caption[]{Schematic representation of the generation of a baroclinic
%flow}
%\end{figure}

\begin{figure}[t]
\centerline{\includegraphics[width=0.6\linewidth]{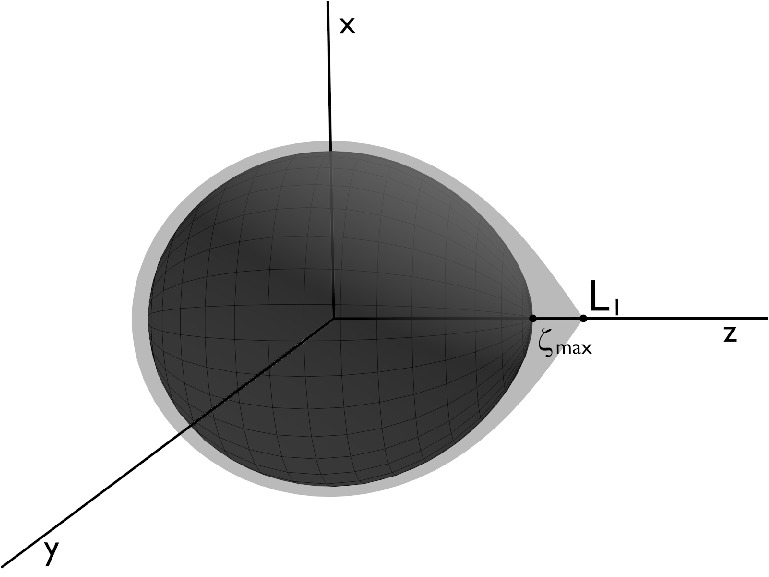}}
\caption{Schematic representation of the primary star with filling
factor $\rho=0.8$. The position of the Lagrange point $L_1$ is shown on
the $z$-axis that joins the centre of the two stars \cite[from][]{ELR12}.}
\end{figure}

\section{Binary stars}\index{binary stars (gravity darkening in)}

Binary stars is another domain where gravity darkening has been
considered, mainly for reproducing the light curves of eclipsing
binaries. We may wonder if the $\omega$-model can be generalized to
predict the gravity darkening of a star belonging to a binary system. It
does but without any (known) analytic solution.

Let us follow the work of \cite{ELR12}. In the radiative envelope of
an early-type star member of a binary system we can still write the
conservation of the flux and assume the anti-parallelism of flux and
effective gravity:

\[ \Div\vF=0 \andet \vF=-f \vg_{\rm eff}\]
but now the effective gravity comes from the 3D potential:

\begin{equation}
\begin{array}{rl}
\phi=&\displaystyle
-\frac{GM_1}{r}-\frac{GM_2}{\sqrt{a^2+r^2-2ar\cos\theta}}\\
&\displaystyle
-\frac{1}{2}\Omega^2r^2(\sin^2\theta\sin^2\varphi+\cos^2\theta)
+a\frac{M_2}{M_1+M_2}\Omega^2r\cos\theta\;,
\end{array}
\end{equation}
where $M_1$ and $M_2$ are the masses of the two stars, 'a' is the
distance between the two stellar centres and $\Omega$ is the orbital
angular velocity. The orbit is assumed circular.
Let us write $\Div(f \vg_{\rm eff})=0$ as

\begin{equation}
\label{eq_grad}
\vec{n}\cdot\nabla\ln
f=\frac{\nabla\cdot\vg_\mathrm{eff}}{g_\mathrm{eff}}\;,
\end{equation}
where we set $\vg_\mathrm{eff}=-g_\mathrm{eff}\vec{n}$.

We consider the three-dimensional curve $\mathcal{C}(\theta_0,\varphi_0)$ that
starts at the centre of the star with the initial direction given
by $(\theta_0,\varphi_0)$, and that is tangent to $\vec n$ at every
point. $\mathcal{C}(\theta_0,\varphi_0)$ is therefore a field line of
the effective gravity field.

The value of $f$ at a point $\vec{r}$ along the curve can be calculated as
a line integral

\begin{equation}
f(\vec r)=f_0\exp\left(\int_{\mathcal{C}(\theta_0,\varphi_0)}
\!\!\!\!\!\frac{\nabla\cdot\vec
g_\mathrm{eff}}{g_\mathrm{eff}}\,\mathrm{d}l\right)
\qquad\mbox{for } \vec r \in \mathcal{C}(\theta_0,\varphi_0)\;.
\label{intf}
\end{equation}

Despite much efforts no analytical expression could be found for $f$.
Expression \eq{intf} is thus integrated numerically.

\begin{figure}[t]
\centerline{\includegraphics[width=0.8\linewidth]{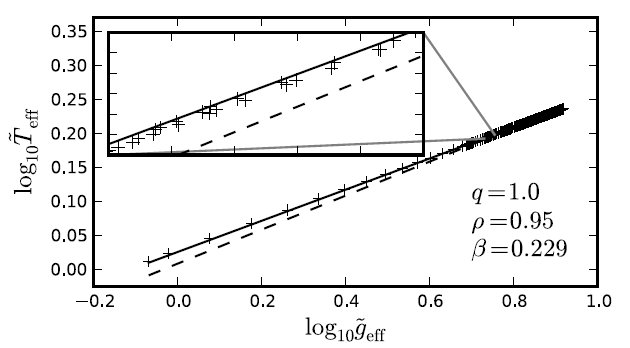}}
\caption[]{From \cite{ELR12}, correlation of effective temperature and
effective gravity in the primary early-type star of a binary system. Mass
ratio is unity and the star fills the Roche lobe at 95\% (see text for
our definition). The correlation may be represented by a $\beta$-exponent
of 0.229. The solid line shows a linear fit, the dashed line the von
Zeipel law, and pluses are from our generalized $\omega$-model.}
\label{tgcorrel}
\end{figure}

One interesting result of this approach, is that there is not a
one-to-one relation between effective gravity and effective temperature.
Indeed, because of the absence of symmetry of the star (except of the
equatorial one if the obliquity is zero), two different points of the
stellar surface may have the same effective gravity but a different
effective temperature. This property comes from expression \eq{intf}:
the path integrals that lead to two points of identical effective
gravity are not necessarily the same and can lead to different values
of the flux. This property is illustrated in Fig.~\ref{tgcorrel}. In
this figure we see that the curve $T_{\rm eff} = f(g_{\rm eff})$ is not
smooth because similar values of $g_{\rm eff}$ lead to different values
of $T_{\rm eff}$. Fortunately, these variations are small.

\begin{figure}
\centerline{\includegraphics[width=0.5\linewidth]{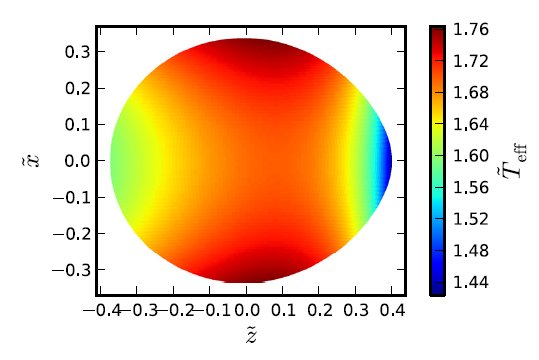}
\includegraphics[width=0.5\linewidth]{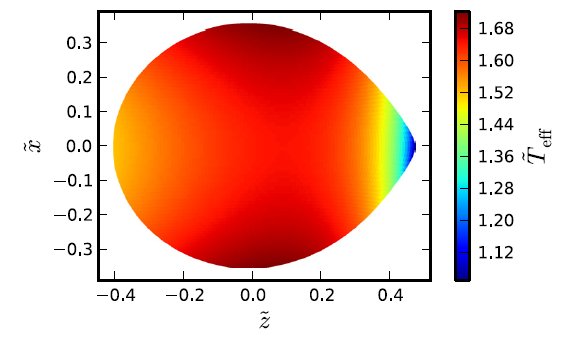}}
\caption[]{From \cite{ELR12}: distribution of the effective
temperature at the surface of a tidally distorted star. Left: $q=1$
and $\rho=0.8$. Right: $q=1$ and $\rho=0.95$.} \label{surf_teff}
\end{figure}

As in \cite{ELR12}, we define $q$ as the mass ratio, and evaluate the
filling of the Roche lobe by the radius $\rho$ of the star along the
line joining the stellar centres, taking the distance between the star
centre and the Lagrange L$_1$ point as unity. Hence, a star filling its
Roche lobe has $\rho=1$ while the one filling it at 95\% has $\rho=0.95$.
The different positions where the same effective temperature are found, is
illustrated in the two cases shown in Fig.~\ref{surf_teff}. There we see
that the curves of isoflux are not simple curves over the stellar surface.

\begin{figure}
\centerline{\includegraphics[height=5cm]{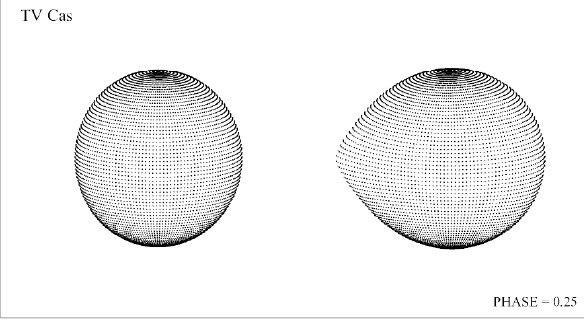}}
\caption{A model of TV Cas that leads to $\beta=0.15$ from fitting the
light curve \cite[from][]{dju03}.}
\label{li0}
\end{figure}

\begin{figure}
\centerline{\includegraphics[height=5cm]{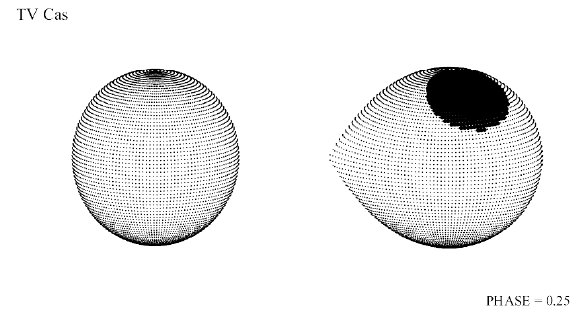}}
\caption{The second model with a spot and $\beta=0.25$ for TV Cas
\cite[from][]{dju03}; the difference between the calculated and observed
light curve is the same as with the model of Fig.~\ref{li1}.}
\label{li1}
\end{figure}

As shown by \cite{dju03}, the determination of the $\beta$-exponent from
the light curves of semi-detached binaries is almost impossible since
magnetic spots induce similar variations (see Fig.~\ref{li0} and
\ref{li1}).

\section{Conclusions}

To conclude these notes, I would like to stress a few points on gravity
darkening:

\begin{itemize}
\item As far as non-magnetic early-type stars are concerned, gravity
darkening has no longer to be proved. The use of the $\beta$-model,
which is not physically sound can be left aside and replaced by the
$\omega$-model, which has the advantage of giving a direct estimate of
the $\omega$ parameter.

\item As far as late-type stars or giant stars are concerned, the
situation is much more uncertain. The problem is indeed more
difficult both on the theoretical and observational sides. On the
theoretical side, the absence of any universally accepted model of
turbulent rotating convection impedes any serious prediction on the
latitude dependence of the convective flux.  Observational constraints
are therefore most welcome.  However, this is not a simple matter
either. Convective envelopes are usually harboring magnetic fields which
can disturb the flux distribution. Ideally, the surface of these stars
should be constrained by both interferometers and Zeeman-Doppler Imaging
so as to disentangle the effects.

\item Finally, for both type of (single) stars, we may recommend the
following scheme of hypothesis and measurements. First assume the axi-
and equatorial symmetry of the star. Then, if the star is centrally
condensed (like an early-type or a giant one), adopt the Roche model. If the
star is not centrally condensed (like a late-type star of the main
sequence), a bipolytropic model is fine. Such a model, which fits the
radiative and convective zones with a polytrope, just depends on three
parameters, mass, equatorial radius and $\omega$, just like the Roche model.
Then, the flux or the effective temperature $T_{\rm eff}(\theta)$ can
be derived after an expansion on the spherical harmonic basis
along with an atmosphere model used for the determination
of the limb darkening effect. The gravity darkening law can then be evaluated
from the curve (or correlation) $T_{\rm eff}(\theta)$ versus $g_{\rm
eff}(\theta)$.

\end{itemize}

\section*{Acknowledgment}
I am grateful to the organizers of the Besan\c con school for their
invitation, and the opportunity to present in more details the recent
work I did with Francisco Espinosa Lara on gravity darkening. This
school triggered many stimulating discussions that helped me deepen
this subject. Finally, I would like to stress that this work owes much
to Francisco who had the original idea of the $\omega$-model.

\section*{Appendix: Angular velocity with respect to critical rotation}
\index{critical rotation}\index{Roche model}

In this appendix we discuss the correspondance between two definitions
of the scaled angular velocity. The first one is the one we used in the
text, namely

\[ \omega = \frac{\Omega}{\Omega_k} =
\Omega\lp\sqrt{\frac{GM}{R_e^3}}\rp^{\!\!-1} \;.\]
where $\Omega_k$ is the orbital angular velocity for an orbit at the actual
equatorial radius of the star.

 The other definition is based on the Roche model and considers the angular
velocity $\Omega_c$ such that the rotation on the equatorial radius is
keplerian. This latter definition is a true critical angular velocity,
while the previous one is a keplerian velocity at the actual equatorial
radius. However, the
critical angular velocity is model dependent, this is why we have to
mention Roche's model for the definition of $\Omega_c$. The first
definition does not need any model, but it is not the exact critical
angular velocity. This latter quantity cannot in general be computed a
priori with just a given spherical model of a star. It needs a full
computation of the structure at the actual critical velocity and is thus
an output of 2D models like ESTER ones \cite[e.g.][]{ELR13}.

So we now only consider Roche models where all quantities can be
derived in a simple manner. We recall that the polar $R_p$ and
equatorial $R_e$ of an equipotential of a star rotating at angular
velocity $\Omega$ are related by

\beq \frac{GM}{R_p} = \frac{GM}{R_e}+\demi \Omega^2R_e^2 \eeq
Then, the critical angular velocity $\Omega_c$ and the critical
equatorial radius $R_{ec}$ are related by

\beq \Omega_c^2 = \frac{GM}{R_{ec}^3}\eeqn{omc}
and hence

\beq R_{ec}=\frac{3}{2}R_p\eeq
at critical rotation.

If the rotation is subcritical, the Roche model gives the following
relation between $R_e$ and $R_p$

\beq R_p=R_e\lp 1+\frac{\omega^2}{2}\rp^{-1}\eeqn{rpre}
But we may take $\Omega_c$ as the scale of the rotation rate and set

\beq \tio = \frac{\Omega}{\Omega_c}\eeq
From the preceding definitions we get the relation between $\tio$ and
$\omega$, namely

\beq \tio=\omega\sqrt{\frac{27}{8}}\lp
1+\omega^2/2\rp^{-3/2}\eeqn{omcomk}
We note that if $\omega=1$ then $\tio=1$ as expected. We also note that
if $\Omega$ is subcritical, then $R_e<3R_p/2$ and therefore
$\Omega_k>\Omega_c$, which implies that we always have

\beq \tio\geq\omega\eeq
Equation \eq{omcomk} shows that it is easy to compute $\tio$ from
$\omega$ but the opposite is a little more complicated since a cubic
equation must be solved. Setting $\chi=\arcsin\tio$, we find

\beq \omega = \sqrt{\frac{6}{\tio}\sin(\chi/3) -2}
\eeq

\bibliographystyle{aa}
\bibliography{../../../biblio/bibnew}

\end{document}